\documentclass[12pt,aps,prl,reprint]{revtex4-1}

\usepackage{color}
\usepackage{graphicx}
\usepackage{epsfig}
\usepackage{epstopdf}
\usepackage{subfig}
\usepackage{amsmath}
\usepackage{bm} 

\def\bB{{\bm B}}

\def\bJ{{\bm J}}

\def\bF{{\bm F}}

\def\br{{\bm r}}
\def\bt{{\bm t}}
\def\bu{{\bm u}}

\def\calO{{\cal O}}

\def\stressT{\stackrel{\leftrightarrow}{\bm{\pi}}}

\def\div{\nabla\cdot}
\def\grad{\nabla}

\def\del{\nabla}
\def\del2{\nabla^2}

\def\curl{\nabla\times}

\def\fsAver#1{\left\langle #1 \right\rangle} 

\def\begeqn{\begin{equation}}
\def\endeqn{\end{equation}}
\def\begeqnar{\begin{eqnarray}}
\def\endeqnar{\end{eqnarray}}
\def\begeqnarn{\begin{eqnarray*}}
\def\endeqnarn{\end{eqnarray*}}

\def\bRhat{\widehat{\bm R}}
\def\bZhat{\widehat{\bm Z}}

\def\PLH{P_{LH}}

\begin{document}
\title[]{Role of edge poloidal density asymmetry in tokamak confinement} 
\author{A. Y. Aydemir}
\email{aydemir@nfri.re.kr}
\affiliation{National Fusion Research Institute, Daejeon 34133, Republic of Korea}                        
\baselineskip 12pt  

\begin{abstract}
Mass flows and radial electric field driven by edge poloidal density asymmetries can be used as a highly effective control mechanism for the edge and thus global confinement in tokamaks. The underlying physics can be demonstrated
entirely within a simple magnetohydrodynamic equilibrium model, without resorting to sophisticated and usually collisionality-dependent neoclassical physics arguments. As an example, strong dependence of the low to high (LH) transition power threshold on the magnetic topology, an experimental observation still poorly understood, can be easily explained  within this framework. Similar arguments indicate that the ITER fueling ports above the midplane might lead to higher input power requirements. 
\end{abstract} 
\maketitle  
 
{\em Introduction--} 
In magnetohydrodynamic (MHD) description of the plasma equilibrium, rapid transport along the field lines leads to a state where the plasma density and temperature are constant on flux surfaces, exhibiting a symmetry in both the poloidal and toroidal directions, the short and long way around the torus, respectively.
This idealization, however,  breaks down at the plasma edge where both the magnetic topology and various perpendicular transport processes introduce at least a poloidal asymmetry.

The first neoclassical studies of the role of poloidally asymmetric transport in the spin-up of a tokamak plasma can be traced back to Stringer\cite{stringer1969}. Others have expanded upon the ``Stringer mechanism'' to explain the origin of the low to high (LH) transition\cite{wagner1982} in terms of the flows generated by inboard-outboard asymmetry of the neoclassical and turbulent particle fluxes\cite{hassam1991, hassam1993}. Similarly, the easier LH transitions observed with inboard gas puffing\cite{valovic2002, field2004} has been explained in terms of neoclassical neutral particle transport
\cite{fulop2002, helander2003, singh2004}. However, in all these, the emphasis has been on the asymmetry of the transport rather than the plasma profiles, and the theoretical treatment has involved intricate neoclassical arguments.
 
There is also a known inverse relationship between plasma dynamics and poloidal asymmetries. Centrifugal forces due to strong toroidal mass flows can introduce an in-out asymmetry in density and pressure profiles\cite{hameiri1983, guazzotto2004}.  Similarly, Alfv\'enic poloidal flows can lead to shocks with sharp poloidal density gradients\cite{shaing1992, seol2016}. Thus, there is an intimate connection between poloidal density or pressure asymmetries and mass flows.

In his work we assume that the nonuniform distribution of fueling sources at the plasma edge, a common feature of present and future tokamaks, can lead to poloidally nonuniform density and pressure profiles. We show that the MHD equilibria consistent with these asymmetries necessarily have strong edge flows and driven radial electric fields. In turn, the flows and fields can have a profound effect on confinement through reduced turbulence\cite{hahm1995} and improved macroscopic stability\cite{bondeson1994, fitzpatrick1996, pustovitov2007b}.

{\em Torque due to an asymmetry--}
If a poloidal asymmetry is introduced into an otherwise symmetric equilibrium,  parallel flows directed away from the local pressure maximum will try to relax the pressure gradient. If the asymmetry is maintained by external sources, as we will assume throughout this work, we expect to see a steady-state poloidal rotation with a definite sign, depending on the net poloidal torque generated. This torque is straightforward to calculate.
 
Consider an axisymmetric plasma boundary curve parametrized as
 \begeqnar
 R(\alpha) & = & R_0 + r\cos{(\alpha + \delta\sin{\alpha})}, \nonumber \\
 Z(\alpha) & = & Z_0 + r\kappa\sin{\alpha}, ~~r=const.,~~0 \le\alpha \le 2\pi, \label{eqn:RZalpha}
 \endeqnar 
 where $\kappa$ is the elongation, and $\delta$ is a measure of the triangularity.
Let $\br(\alpha) \equiv [R(\alpha)-R_0]\bRhat + [Z(\alpha)-Z_0]\bZhat$, and define the tangent vector $\bt \equiv \partial\br/\partial\alpha.$
The force in the direction of $\bt$ due to the pressure gradient is $
\bF_\alpha = -(1/h_\alpha^2)\bt(\bt\cdot\grad)p,$ 
where $h_\alpha \equiv \sqrt{\partial\br/\partial\alpha\cdot \partial\br/\partial\alpha}$. Then the torque $T_\zeta  =  (\br\times\bF_\alpha)_\zeta$ is given by
\begeqn
T_\zeta  =  -\frac{1}{h_\alpha^2}\left\{(R-R_0)\frac{\partial Z}{\partial\alpha} - (Z-Z_0)\frac{\partial R}{\partial\alpha}\right\}\frac{\partial p}{\partial\alpha}. \label{eqn:Torque0}
\endeqn
Here $\zeta$ is the usual toroidal angle of the $(R,Z,\zeta)$ cylindrical coordinates.
The net torque is given by the surface-average of $T_\zeta$, defined as $\fsAver{T_\zeta}  \equiv  (1/K)\oint T_\zeta Rh_\alpha  d\alpha d\zeta,$  where $K \equiv \oint Rh_\alpha d\alpha d\zeta.$ 
Assuming the pressure asymmetry is due to a perturbation with a ``wrapped Gaussian'' profile centered at $\alpha=\alpha_0$, we have $p(r,\alpha) = p_0(r) + \delta p(r,\alpha;\alpha_0)$ and
\begeqn
\delta p(r,\alpha; \alpha_0) = \delta p(r) \sum_{k=-\infty}^\infty e^{-(\alpha-\alpha_0 + 2\pi k)^2/w^2}.\label{eqn:pProfile}
\endeqn
We can easily evaluate the surface-average of $T_\zeta$ in circular geometry. Setting $\delta=0$, $\kappa=1$ leads to $h_\alpha=r,~T_\zeta  =  -\partial p/\partial \alpha$, and
\begeqn
\fsAver{T_\zeta}  =  -\frac{r\delta p(r)}{2\pi R_0}\sum_{k=-\infty}^\infty  \int_0^{2\pi} e^{-(\alpha-\alpha_0 + 2\pi k)^2/w^2} \sin\alpha d\alpha. \label{eqn:averTorque}
\endeqn
For $w \ll 2\pi$, $\fsAver{T_\zeta}$ as a function of $\alpha_0$ has a sinusoidal form. For $\delta p > 0,$ it is positive (negative) if the pressure nonuniformity is in the lower (upper) half plane.

\begin{figure}[htbp]
\begin{center}
\includegraphics[width=3in]{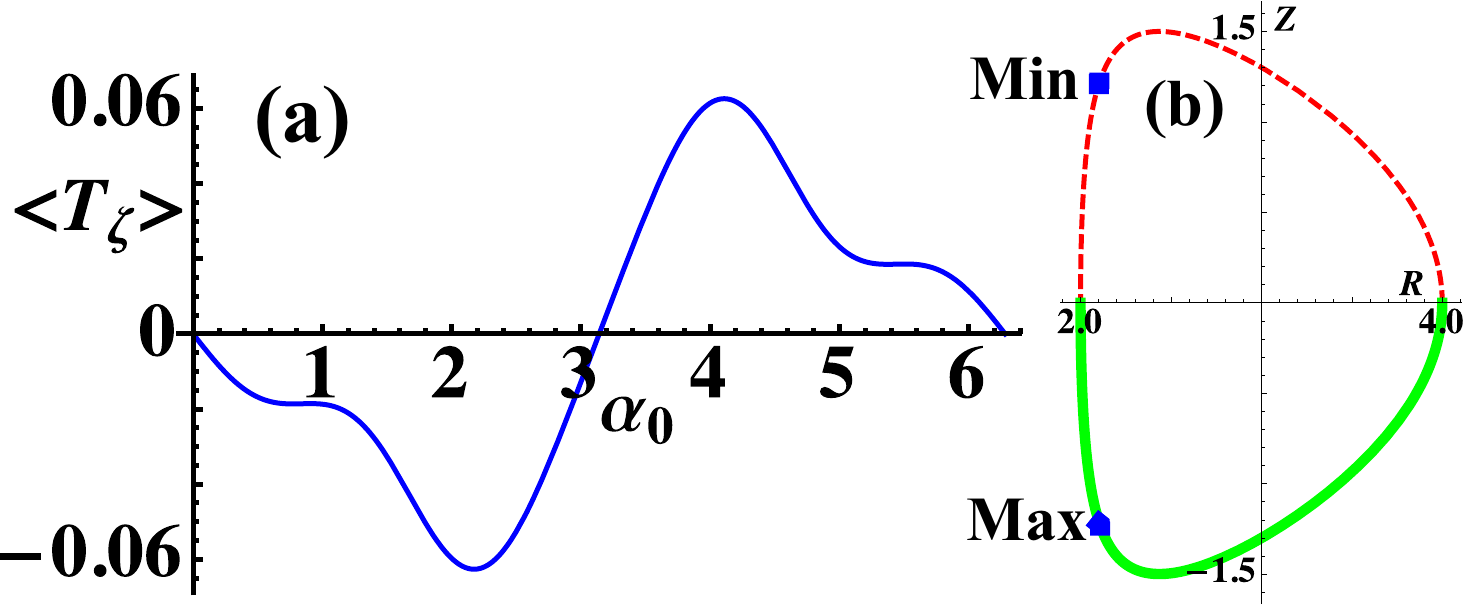}
\caption{\baselineskip 12pt (a) Surface-averaged torque $\fsAver{T_\zeta}$ as a function of $\alpha_0$ (see Eq.~\ref{eqn:pProfile}) for $a/R_0=1/3,~Z_0=0,~\kappa=1.5$, $\delta=0.6,~\delta p=1,~w=\pi/4.$ (b) The poloidal geometry. A density nonuniformity in the region shown in solid green will lead to a negative (favorable) radial electric field. Locations of the minimum and maximum torque are marked.}
\label{fig:torqueDN}
\end{center}
\end{figure}

For a more general geometry the net torque $\fsAver{T_\zeta}$ can be evaluated numerically. An example is shown in Fig.~\ref{fig:torqueDN} for $\delta=0.6,~\kappa=1.5.$  Again, $\fsAver{T_\zeta}$ is positive for a perturbation in the lower half-plane. 
For the ``standard configuration'' of toroidal field and current (both clockwise when seen from above), a positive torque will drive a positive (in the electron diamagnetic drift direction) poloidal rotation, which in turn will make a negative contribution to the edge radial electric field. Note that the minimum and maximum torque occur near where  the upper and lower X-points would be, respectively, for a diverted tokamak of this poloidal geometry. Thus,  the largest positive poloidal flow and most negative radial electric field would be generated for $\alpha_0=4.1$, near the lower X-point. For this up-down symmetric system, the torque vanishes at both the outer ($\alpha_0=0$) and inner ($\alpha_0=\pi$) midplane. This is no longer true when the up-down symmetry is broken.

\begin{figure}[htbp]
\begin{center}
\includegraphics[width=3.3in]{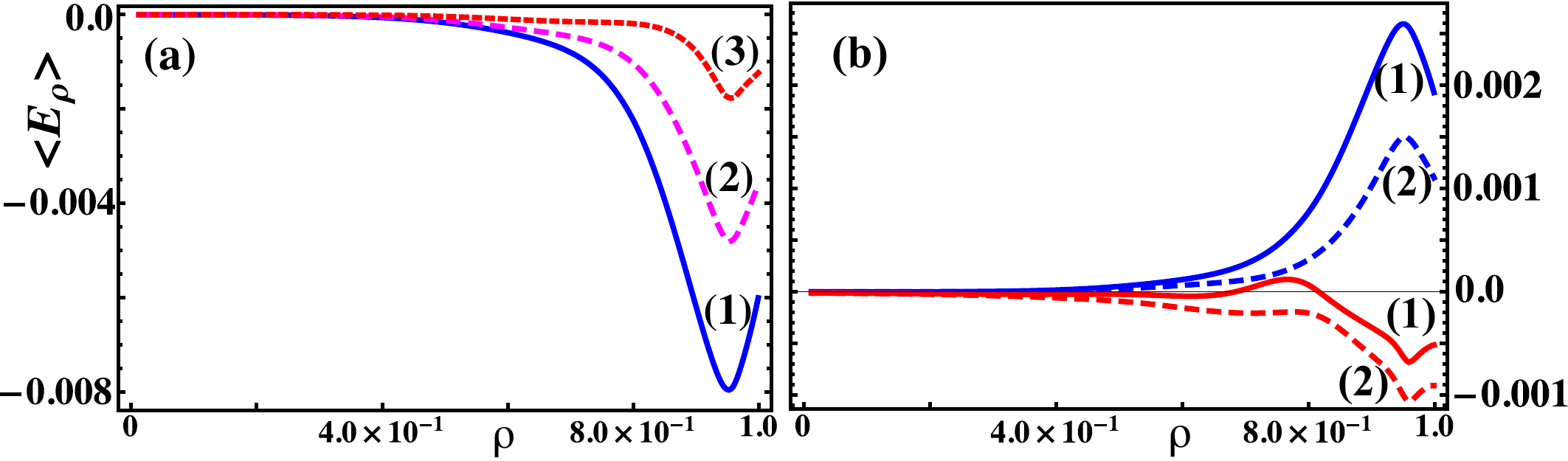} 
\caption{\baselineskip 12pt (a) $\fsAver{E_\rho}$ in steady-state as a function of radius for $\omega_0=4.28$.  (b) $\fsAver{u_\omega}$ (blue) and $\fsAver{u_\zeta}$ (red). For both panels, the poloidal flow damping rates are: (1) $\gamma_p=0$, (2) $\gamma_p=10^{-4}$, (3) $\gamma_p=10^{-3}$.}
\label{fig:summaryB}
\end{center}
\end{figure}

{\em Asymmetric equilibrium calculations with CTD--}
Equilibrium calculations with poloidally nonuniform density profiles start with a static MHD equilibrium with $p=p(\psi)$, which is then gradually modified with a perturbation $\delta p=T(\psi)\delta\rho_m$ of the form shown in Eq.~\ref{eqn:pProfile}. Because of large parallel thermal conductivity, the temperature is assumed to be a flux function. A neighboring equilibrium with mass flows but without poloidal symmetry is obtained as the time-asymptotic limit of an initial-value calculation with the CTD code (see \cite{aydemir2015} and the references therein). Typically $\delta \rho_m/\rho_{edge} \sim  \calO(10^{-1})$, where $\rho_{edge}$ is the edge mass density for the initial equilibrium.
The momentum equation has the form $\rho_md\bu/dt=\bJ\times\bB - \grad p -\div\stressT - \gamma_p \rho_m\bu_p,$ where $\div\stressT = \curl \mu \curl\bu - (4/3)\grad\mu\div\bu$ is an isotropic viscosity term. The last term represents poloidal flow damping\cite{hassam1978}. It is applied only to the poloidal flow. For the damping rate we use an expression given by Shaing\cite{shaing2015}:  $\gamma_p = 0.68\nu_{ii}/\epsilon$, where $\nu_{ii}$ is the ion collision frequency and $\epsilon$ is the inverse aspect ratio.
\begin{figure}[htbp]
\begin{center}
\includegraphics[width=2.5in]{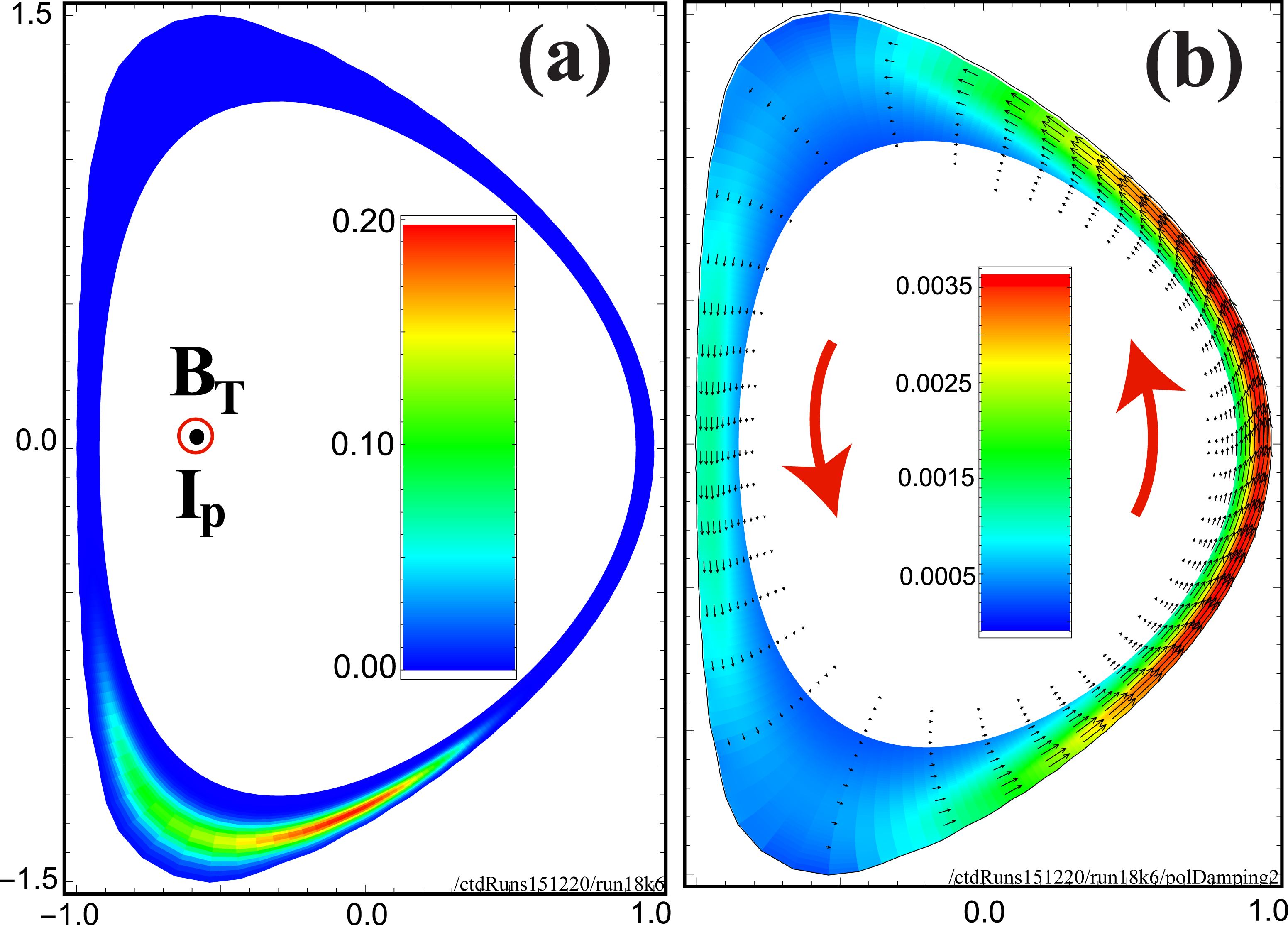} 
\caption{\baselineskip 12pt (a) Density perturbation at $\omega_0=4.28$ with an amplitude $\delta\rho_m=0.2$ and poloidal width $w_\omega=0.3.$ $\gamma_p=10^{-4}.$  (b) Poloidal velocity vectors, all in the $+\omega$ direction. Only the edge is shown.}
\label{fig:summaryA}
\end{center}
\end{figure}
The CTD code uses a conformal, orthogonal $(\rho,\omega,\zeta)$ coordinate system, where $\rho, \omega$ are the radial and  poloidal  coordinates, respectively, and $\zeta$ is the usual toroidal angle\cite{aydemir1990}. A density perturbation $\delta \rho_m(\rho,\omega)$ in the lower half-plane and the resulting positive poloidal flow are shown in Fig.~\ref{fig:summaryA}. Here $\delta \rho_m(\rho,\omega; \omega_0)$ has the same functional form as $\delta p(r,\alpha; \alpha_0)$ in Eq.~\ref{eqn:pProfile}, with $\delta\rho_m(\rho)$ given also by a Gaussian. The radial width is $w_\rho=0.025$ throughout this work. The  width of the wrapped poloidal Gaussian is $w_\omega =  0.3$ in Fig.~\ref{fig:summaryA} (a). 
Details of the radial electric field and rotation velocities for the new equilibrium are shown in Fig.~\ref{fig:summaryB}. In (a), the surface-averaged $E_\rho$ as a function of the radial coordinate is plotted for three different values of the poloidal flow damping rate. The panel (b) shows the surface averages for the poloidal (blue) and toroidal (red) velocities for two different damping rates. Note that with finite damping rate ($\gamma_p=10^{-4}$), the poloidal flow decreases, as expected, but the toroidal flow increases. This trend continues for $\gamma_p=10^{-3}$.
The dimensionless viscosity coefficient  is $\mu = 10^{-6}$ for all the cases shown.
Note the prominent negative radial electric field well centered around the radial location of the poloidal nonuniformity in Fig.~\ref{fig:summaryB} (a). This field is driven largely by the poloidal rotation for $\gamma_p=0$, but the negative (counter-current) toroidal rotation makes an increasing contribution with finite damping rate.
 
\begin{figure}[htbp]
\begin{center}
\includegraphics[width=3.25in]{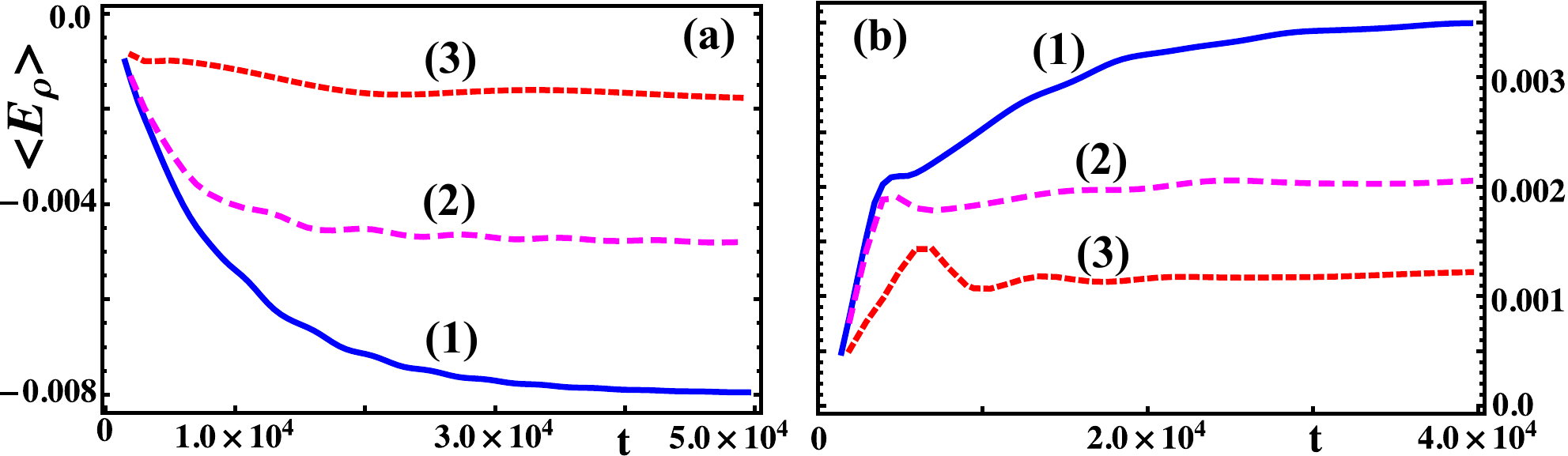} 
\caption{\baselineskip 12pt  Time histories for $\fsAver{E_\rho}$ for two poloidal locations: (a) $\omega_0=4.8$. (b) $\omega_0=1.2$. The poloidal flow damping rates: (1) $\gamma_p=0$, (2) $\gamma_p=10^{-4}$, (3) $\gamma_p=10^{-3}$.}
\label{fig:ErhoHistories}
\end{center}
\end{figure}

The computational results are shown in non-dimensional form. The velocities have been normalized to the poloidal Alfv\'en speed defined in terms of the central toroidal field strength and edge mass density: $v_{Ap} \equiv \epsilon B_{\zeta 0}/\sqrt{\mu_0\rho_{edge}},$  where $\epsilon \equiv a/R_0.$ Using $B_{\zeta 0}=3T,~n_{edge}=10^{19} m^{-3},a=1m,~~\epsilon=1/3$ and a deuterium plasma, we get
$v_{Ap} = 4.9\times 10^6 ms^{-1}$, which gives the poloidal Alfv\'en time $\tau_{Ap} = 2.0\times 10^{-7} s.$ For these parameters, the poloidal damping rate\cite{shaing2015} is $\gamma_p=3.3\times 10^2 s^{-1}.$ Normalizing to $\tau_{Ap}$ yields $\gamma_p=6.7\times 10^{-5}.$ Thus the damping rate $\gamma_p=10^{-4}$ used in some of the CTD calculations is somewhat higher than implied by the physical parameters assumed; the value $\gamma_p=10^{-3}$ is more than an order of magnitude higher. In dimensional units the viscosity coefficient $\mu=10^{-6}$ corresponds to a momentum diffusivity of $\chi_\varphi=5m^2 s^{-1}$, which is not unrealistic. A consistent normalization factor for the electric field is $E_0=\epsilon v_{Ap}B_{\zeta 0} = 4.9\times 10^6 Vm^{-1}$,
which leads to $\fsAver{E_\rho}_{min}~=~-24 kVm^{-1},~8.7keV$ for $\gamma_p=10^{-4},~10^{-3}$, respectively. Both represent substantial fields. Recall, however, we assumed $\delta\rho_m/\rho_{edge}=0.2$ above. Since the torque is linear in the perturbation amplitude (see Eq.~\ref{eqn:averTorque}), and since this quantity is not easily available, our numerical results should be interpreted in the light of this uncertainty. 
\begin{figure}[htbp]
\begin{center}
\includegraphics[width=2.5in]{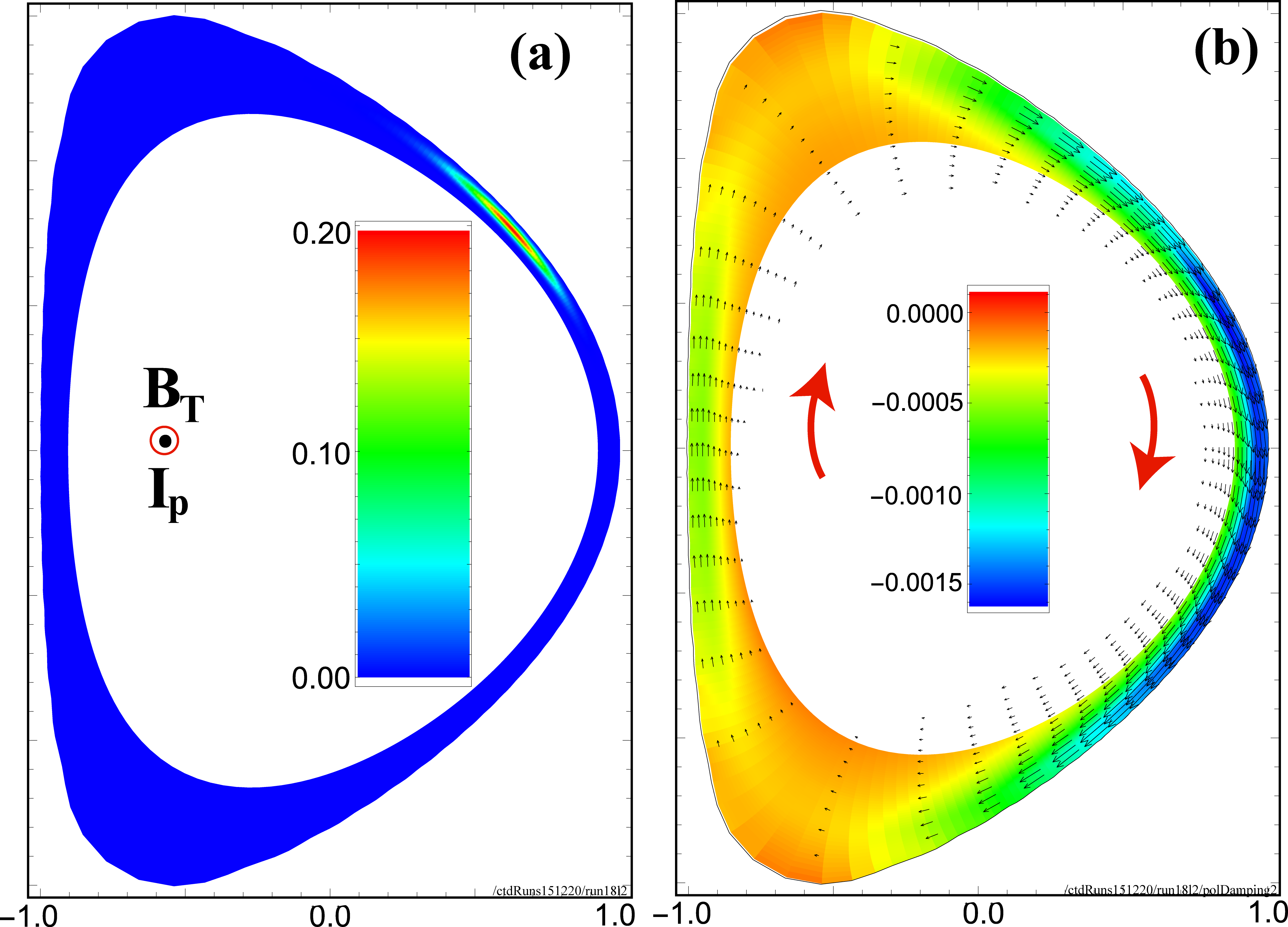} 
\caption{\baselineskip 12pt (a) Density perturbation at $\omega_0=1.2$. (b) The poloidal velocity is uniformly negative.}
\label{fig:summaryA_run18l2}
\end{center}
\end{figure}
As expected from our discussion on the net torque (see Fig.~\ref{fig:torqueDN}), for the same magnetic geometry, a density nonuniformity in the upper half-plane  as in Fig.~\ref{fig:summaryA_run18l2} (a) ($\omega_0=1.2$) reverses the poloidal flow (panel (b)) and the radial electric field. Time histories of the electric field for these two cases are shown in Fig.~\ref{fig:ErhoHistories} for three different values of the damping rate $\gamma_p.$ Note that all have reached quasi-steady-states at the end of the calculations. 

\begin{figure}[htbp]
\begin{center}
\includegraphics[width=2.5in]{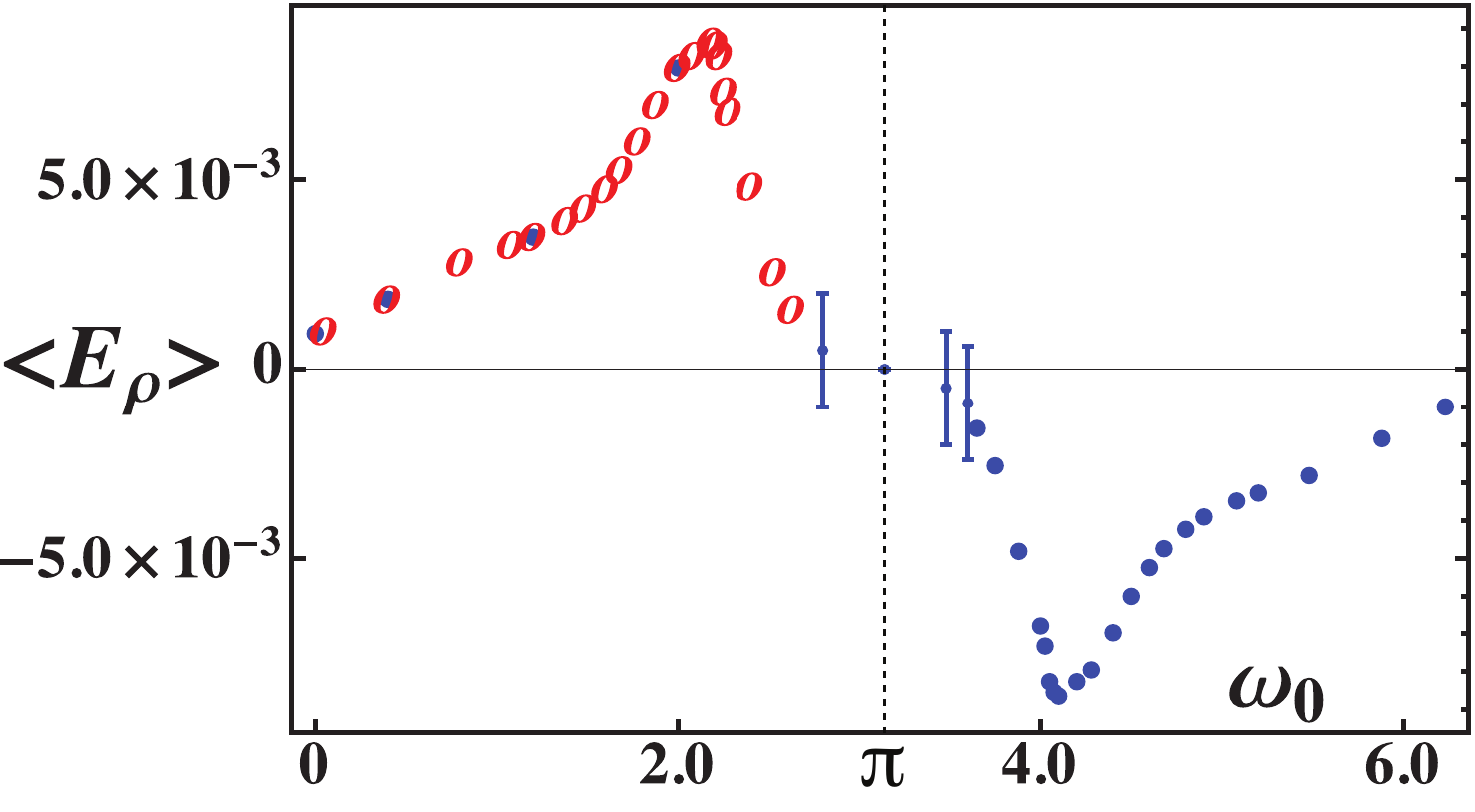}
\caption{\baselineskip 12pt $\fsAver{E_\rho}$ as a function of the poloidal location $\omega_0$ of the density perturbation. For $\omega_0\simeq \pi$, no steady-state solution exists; the ``error bars'' show the range of oscillations observed. Here $\gamma_p=0$ has been assumed; finite $\gamma_p$ is expected to change the vertical scale without qualitatively altering the results.}
\label{fig:ErhoScan}
\end{center}
\end{figure}

The radial electric field strongly depends on the poloidal location of the density perturbation, as seen in Fig.~\ref{fig:ErhoScan}. This variation with the angle $\omega_0$ is in good agreement with the surface-averaged torque of Fig.~\ref{fig:torqueDN}. (Recall that positive torque leads to positive poloidal rotation and negative electric field.) The electric field minimum is at $\omega_0 = 4.08$, which is consistent with the location of the torque maximum in Fig.~\ref{fig:torqueDN} (b). Note that this point is to the left (high-field side) of where we would expect the X-point to be located if this were a diverted geometry (see also Fig.~\ref{fig:torqueDN} (b)). In the figure only the solid (blue) circles represent actual calculations with CTD. The open (red) ovals have been obtained  using odd symmetry about the midplane implied by the net torque of Fig.~\ref{fig:torqueDN}: $\fsAver{E_\rho}(\omega_0)=-\fsAver{E_\rho}(2\pi-\omega_0).$ This symmetry is confirmed by actual calculations at a few locations above the midplane (solid blue circles).
Solutions near the inner midplane, $\omega_0\simeq \pi,$ are oscillatory in time. The range of oscillations in the electric field is shown with error bars in Fig.~\ref{fig:ErhoScan} for the points at $\omega_0=2.80,~\omega_0=2\pi-2.80$, and $\omega_0=3.60$. The period of the oscillations is consistent with sound waves with a frequency $\omega_s =q_eR/c_s,~c_s=\sqrt{\gamma p/\rho_m}=\sqrt{T},$ using $\gamma=1,~p=\rho_m T,~q_e \simeq 5.$

From the arguments leading to Eq.~\ref{eqn:averTorque} and Fig.~\ref{fig:torqueDN}, it is clear that the sign of the net torque is determined only by the toroidal geometry and not by details of the fields. However, with the perturbation location fixed, sign of the toroidal flow and radial electric field depend on the direction of the toroidal and poloidal fields. General symmetry arguments show that the MHD equations remain invariant under reversal of the toroidal field or current ($B_\zeta \rightarrow -B_\zeta,$ or $J_\zeta \rightarrow -J_\zeta$, but not both) only if the toroidal flow also reverses, $u_\zeta \rightarrow -u_\zeta$\cite{aydemir2007b}. Thus, the radial electric field $E_\rho = -u_\omega B_\zeta + u_\zeta B_\omega$ reverses with the toroidal field, but not with the toroidal current. As a result, the favorable (solid green) and unfavorable (dashed red) regions are interchanged in Fig.~\ref{fig:torqueDN} (b) with $B_\zeta \rightarrow -B_\zeta$. 

\begin{figure}[htbp]
\begin{center}
\includegraphics[width=2.75in]{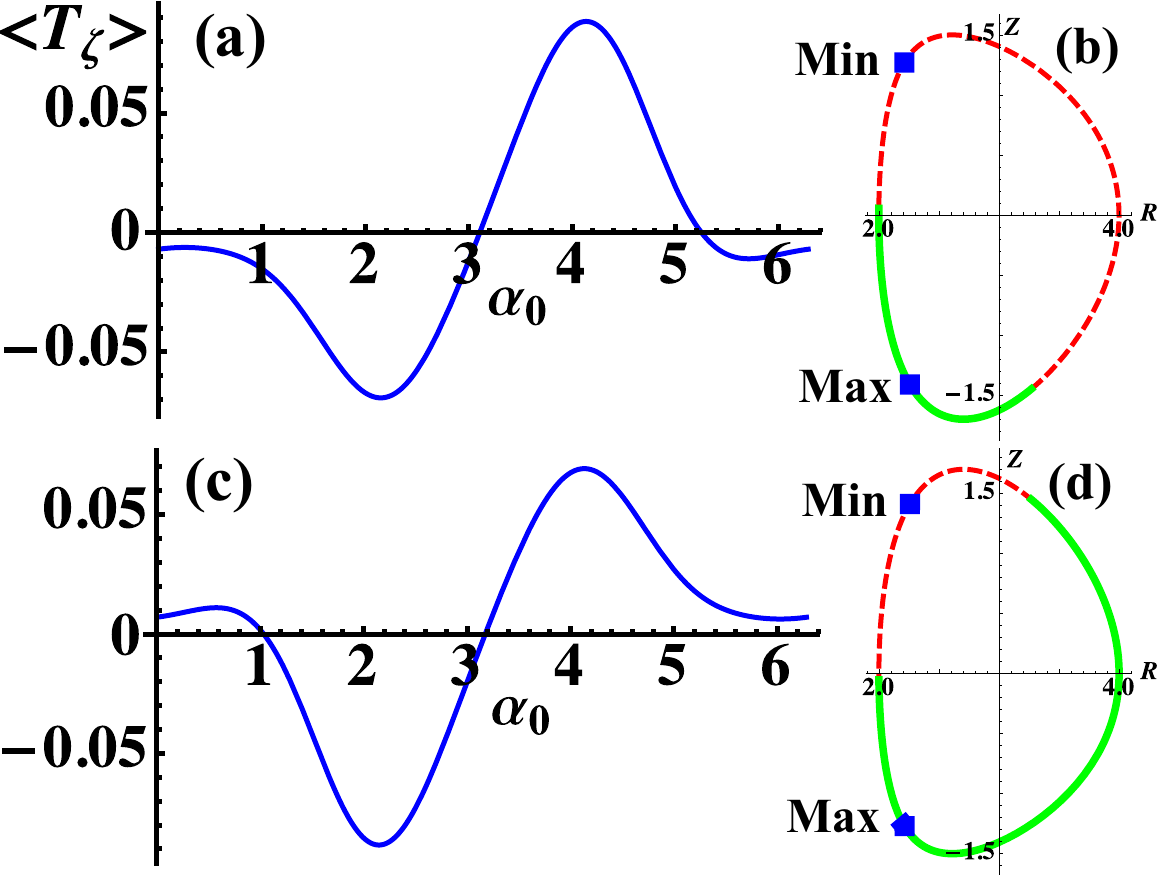}
\caption{\baselineskip 12pt (a) Torque $\fsAver{T_\zeta}$ as a function of $\alpha_0$ for a LSN-like magnetic geometry with $\kappa_U=1.5$, $\kappa_L=1.7$, $\delta_U=0.4$, $\delta_L=0.3$, $\delta p=1$, $w=\pi/4$. (b) The poloidal geometry. The panels (c) and (d) are for a correspoinding USN geometry. A perturbation in regions shown in solid green will generate a negative (favorable) radial electric field.}
\label{fig:torqueCombined}
\end{center}
\end{figure}

{\em Up-down asymmetric geometries--}
The odd sinusoidal symmetry of the net torque in the symmetric double-null (DN)-like field configuration of Fig.~\ref{fig:torqueDN} (b) is broken in more general geometries. This point is demonstrated in Fig.~\ref{fig:torqueCombined} for  lower and upper-single-null (LSN, USN)-like magnetic geometries. Here the elongation and triangularity of Eq.~\ref{eqn:RZalpha} are defined in terms of their upper and lower parts as $\kappa \equiv \left[(\kappa_U+\kappa_L) + (\kappa_U-\kappa_L)\sin\alpha\right]/2$, and $\delta \equiv \left[(\delta_U+\delta_L) + (\delta_U-\delta_L)\sin\alpha\right]/2$. The LSN case (panels (a) and (b)) have $\kappa_U=1.5,~\kappa_L=1.7$, and $\delta_U=0.4,~\delta_L=0.3.$ The USN geometry of panels (c) and (d) have these values interchanged.

For the LSN geometry,  (Figs.~\ref{fig:torqueCombined} (a) (b)), the loss of up-down symmetry shrinks the portion of the boundary favorable for poloidal density nonuniformity essentially to the lower left quadrant: a perturbation in the region shown in solid green in Fig.~\ref{fig:torqueCombined} (b) would lead to a positive poloidal flow and a negative radial electric field.  All other regions (dashed red line) would generate a negative poloidal flow and positive electric field. Note that the outer midplane and surroundings ($\alpha_0\simeq 0$) are now entirely in the unfavorable region. Switching from LSN to USN (panels (c), (d)) exchanges the favorable and unfavorable regions. Now only the upper left quadrant is undesirable for a poloidal density perturbation. 

{\em Discussion and summary--}
Poloidal density asymmetries and their poloidal location in an axisymmetric MHD equilibrium strongly influence both the direction and amplitude of the generated flows and radial electric field at the tokamak edge. Since these are known to play an important role in confinement and stability, a deliberately introduced asymmetry can be used as an effective control knob for the LH transition, edge localized mode (ELM) mitigation and other problems. As an example, we will make use of the naturally occurring density asymmetries near the X-points of a diverted tokamak to explain the magnetic geometry dependence of the LH transition power threshold, $\PLH$: When the  ion $\grad B$ drift is in the direction of the active X-point (favorable direction), $P_{LH}$ is about a factor of two lower than when it is in the unfavorable direction (away from the active X-point)\cite{ryter1996, carlstrom2005}. 

Neutral recycling near the X-points is a major source of fueling\cite{fukuda2000, groth2011}, which results in a poloidally localized density and pressure increase inside the separatrix near the same area\cite{carreras1998}. For a double-null (DN) configuration, these regions approximately correspond to the points labelled ``Min'' and ``Max'' in Fig.~\ref{fig:torqueDN} (b). Thus, in a balanced DN, the effects of the pressure asymmetries at the top and bottom will cancel each other, and there will be no net poloidal rotation or radial electric field generated (due to the asymmetries). For the LSN geometry of Fig.~\ref{fig:torqueCombined} (b), however, recycling near the point labelled ``Max'' will lead to a negative $\fsAver{E_\rho}$ and reduce $P_{LH}$, since the H-mode is associated with a deepening negative radial electric field well\cite{shaing1989, carlstrom2005}. For the USN in Fig.~\ref{fig:torqueCombined} (d),  recycling near the point ``Min'' will make a positive contribution to the radial electric field and increase $P_{LH}$. Thus we can conclude that $P_{LH}^{LSN} < P_{LH}^{DN} < P_{LH}^{USN}$, which has strong experimental support\cite{ryter1996}. Here we have assumed the fields are in the ``standard configuration.'' If the toroidal field is reversed without changing the poloidal configuration, the resulting reversal of the electric field will reverse the inequality signs and lead to 
$P_{LH}^{LSN} > P_{LH}^{DN} > P_{LH}^{USN}.$ These results are consistent with the increase in $P_{LH}$ when the ion $\grad B$ drift direction points away from the active X-point. Note that the arguments above are solely based on the effects of naturally occurring neutral recycling near the X-point. If an asymmetry is deliberately introduced, for example, near the point labeled ``Max'' in the USN geometry of Fig.~\ref{fig:torqueCombined} (d), with the unfavorable drift direction (down), it will {\em reduce} the power threshold $\PLH$, an effect that can be easily tested experimentally.

Our results may have some negative implications for the ITER fueling system\cite{baylor2007}. Our findings above imply that the gas injectors above the midplane will drive a negative poloidal flow and a positive radial electric field and thus increase $\PLH$. In contrast, the gas injectors near the divertor are optimally located.

In summary,  a localized fueling source, a common feature of present and future devices,  can be used as a highly effective external knob to control the edge confinement and thus the global confinement in tokamaks. In particular, the flows and radial electric field driven by poloidal density asymmetries near the X-points can provide a simple and robust explanation for the widely-observed dependence of the LH transition power threshold on magnetic topology.

\begin{acknowledgments}This work was supported by MSIP, the Korean Ministry of Science, ICT and Future Planning, through the KSTAR project.
\end{acknowledgments}


\end{document}